\documentclass[a4paper]{jpconf}
\usepackage{graphicx}
\usepackage{siunitx}
\usepackage[square,sort&compress,comma,numbers]{natbib}
\usepackage{amsmath,amssymb}
\usepackage{floatrow}
\usepackage{url} %fixes use of special characters in url 
\usepackage[export]{adjustbox}

\newfloatcommand{capbtabbox}{table}[][\FBwidth]
\DeclareSIUnit\gauss{G}

\begin{document}
\title{A cold electron-impact ion source driven by a photo-cathode – New opportunities for the delivery of radioactive molecular beams?}

\author{Jochen Ballof\textsuperscript{1,2}, Mia Au\textsuperscript{1,2}, Ermanno Barbero\textsuperscript{1}, Katerina Chrysalidis\textsuperscript{1}, Christoph E Düllmann\textsuperscript{2,3,4}, Valentin Fedosseev\textsuperscript{1}, Eduardo Granados\textsuperscript{1}, Reinhard Heinke\textsuperscript{1}, Bruce A Marsh\textsuperscript{1}, Michael Owen\textsuperscript{1}, Sebastian Rothe\textsuperscript{1}, Thierry Stora\textsuperscript{1} and Alexander Yakushev\textsuperscript{3,4}}

\address{\textsuperscript{1} CERN, Accelerator Systems Department, 1211 Geneva 23, Switzerland}
\address{\textsuperscript{2} Johannes Gutenberg - Universit\"{a}t Mainz, Department Chemie, Standort TRIGA, Fritz-Strassmann-Weg 2, 55128 Mainz, Germany}
\address{\textsuperscript{3} GSI Helmholtzzentrum f\"{u}r Schwerionenforschung, 64291, Darmstadt, Germany}
\address{\textsuperscript{4} Helmholtz-Institut Mainz, 55099 Mainz, Germany}

\ead{jochen.ballof@cern.ch, thierry.stora@cern.ch}

\begin{abstract}
The thick-target ISOL (Isotope mass Separation OnLine) method provides beams of more than 1000 radionuclides of 74 elements. The method is well established for elements with sufficiently high volatility at ca. \SI{2000}{\celsius}. To extract non-volatile elements the formation of a volatile molecule is required. While successful in some cases (e.g. carbon or boron), most of these elements are not yet available as ISOL beam. A variety of volatile carrier molecules has been proposed for all elements produced in the target material, but their probability of survival during the extraction and ionization process is often limited by the high temperatures required for isotope diffusion in the thick targets and for ion source operation. While cold target concepts have already been proposed, the normal mode of operation of the typically used Versatile Arc Discharge Ion Source (VADIS) with a hot cathode is not well suited. Here, we report about first measurements with an electron-impact ion source operated at ambient temperature using electrons that were liberated via the photo-electric effect from a copper cathode. 

\end{abstract}

\section{Introduction}

Since the advent of the ISOL (Isotope Separation On-Line) method in the 1960s, facilities like CERN-ISOLDE \cite{Borge2017,Catherall2017} have been erected, which provide beams of a variety of radioisotopes. To keep up with the demand for ever more exotic and intense radioactive ion beams, extensive upgrade programs are on the way, \textit{e.g.}, at GANIL (Caen, France) or TRIUMF (Vancouver, Canada) \cite{Blumenfeld2013}. The ISOL technique exploits thick targets (typically at least tens of grams per \si{\centi\meter\squared}) in which nuclides are produced in reactions induced by an energetic driver beam. The nuclides are stopped within the target material and have to diffuse out of the target matrix and effuse into the ion source from where they are electrostatically extracted and separated by mass-to-charge ratio. The target is typically kept at high temperatures of up to ca. \SI{2000}{\celsius} to promote diffusion and effusion processes. Even at these elevated temperatures, many elements cannot be extracted due to their refractory properties, \textit{i.e.}, high melting and boiling points, which prohibit their transport. 

A powerful method to volatilize elements is the \textit{in-situ} formation of a volatile carrier molecule \cite{Sidenius1961}.  While volatile carrier molecules have been proposed for almost all elements \cite{Freeman1973,Herrmann1967}, the \textit{in-situ} volatilization for radioactive ion beam formation was so far only successful for few refractory elements like carbon \cite{franberg2008production} or boron, \cite{Ballof2019} which form compounds that are stable in a high-temperature environment \cite{KoesterImpossibleBeams,Kronenberg2008}.  Suitable compound classes are mostly restricted to binary halides or chalcogenides. To increase the variety of volatile carrier compounds, the temperatures of the target and ion source need to be significantly reduced. Cold target concepts have been proposed, which exploit the fast diffusion in nano-materials \cite{RAMOS201681} or fully avoid diffusion and exploit the recoil momentum of nuclear reaction products \cite{ballof2021concept}. However, the ion sources typically used at ISOLDE are not well suited for the ionization of more delicate volatile compounds \cite{Jochendiss}. 

A review about ion sources for radioactive ion beams can be found, \textit{e.g.}, in ref.~\cite{Stora:1693046}. For compounds with low ionization potential ($\lesssim \SI{6}{\electronvolt}$), hot cavities facilitating surface ionization are typically used. While most elements with higher ionization potential can be efficiently ionized by element-selective resonant laser ionization \cite{Marsh:1967371}, the ionization of compounds is addressed with electron impact ionization. Typically, the Versatile Arc Discharge Ion Source (VADIS) \cite{Penescu2010,LiviuDiss} is used for such compounds at ISOLDE. For surface ionization, the efficiency directly depends on the temperature according to the Saha-Langmuir law \cite{Langmuir1925}. In the VADIS, electrons are released thermionically according to the  Richardson–Dushman equation \cite{Wolf2017}. This equally requires high temperatures. Radio-frequency heated plasma ion sources have been proposed \cite{Gaubert2003,Kronberger_2013,suominen2010ionization} which, however, suffer from instabilities in pulsed primary beam operation \cite{Jardin2003} and an electron energy distribution that often favors breakup of the molecule over ionization \cite{PekkaTalkARIS,Jochendiss}.

We investigated the operation of the VADIS at ambient temperature by exploiting laser-induced electron release. We have previously reported on a VADIS operated at ambient temperature in which electrons were released from the tantalum cathode by a femto-second laser \cite{ballof2021concept}. We were able to show that the fragile carrier molecule Mo(CO)\textsubscript{6} can be ionized with efficiencies in the same order of magnitude as the noble gas krypton in the same setup. A parameter study was presented that indicates that an ion source based on electron production via the photo-electric effect could reach an efficiency of ca. \SI{1}{\percent}. However, the mechanism of electron production was not  identified in our previous experiment.

In this work, we have tailored our proof-of-concept experiment to the release of electrons via the photo-electric effect. A copper cathode was used which exhibits a relatively high quantum efficiency and does not require extreme vacuum conditions \cite{Pimpec2013}. The wavelength, pulse energy and pulse length of the laser were chosen such that ablation and thermionic electron generation could be excluded. Non-volatile fragments can be produced in electron-molecule collisions and condense on the surface of a photo-cathode. This work provides a first insight into the influence of these deposited fragments on the quantum efficiency of a photo-cathode.

\floatsetup[table]{capposition=top}

\begin{figure}
\CenterFloatBoxes
\begin{floatrow}
\ffigbox[22pc]{%
  %\rule{20pc}{3cm}%
  \includegraphics[width=22pc]{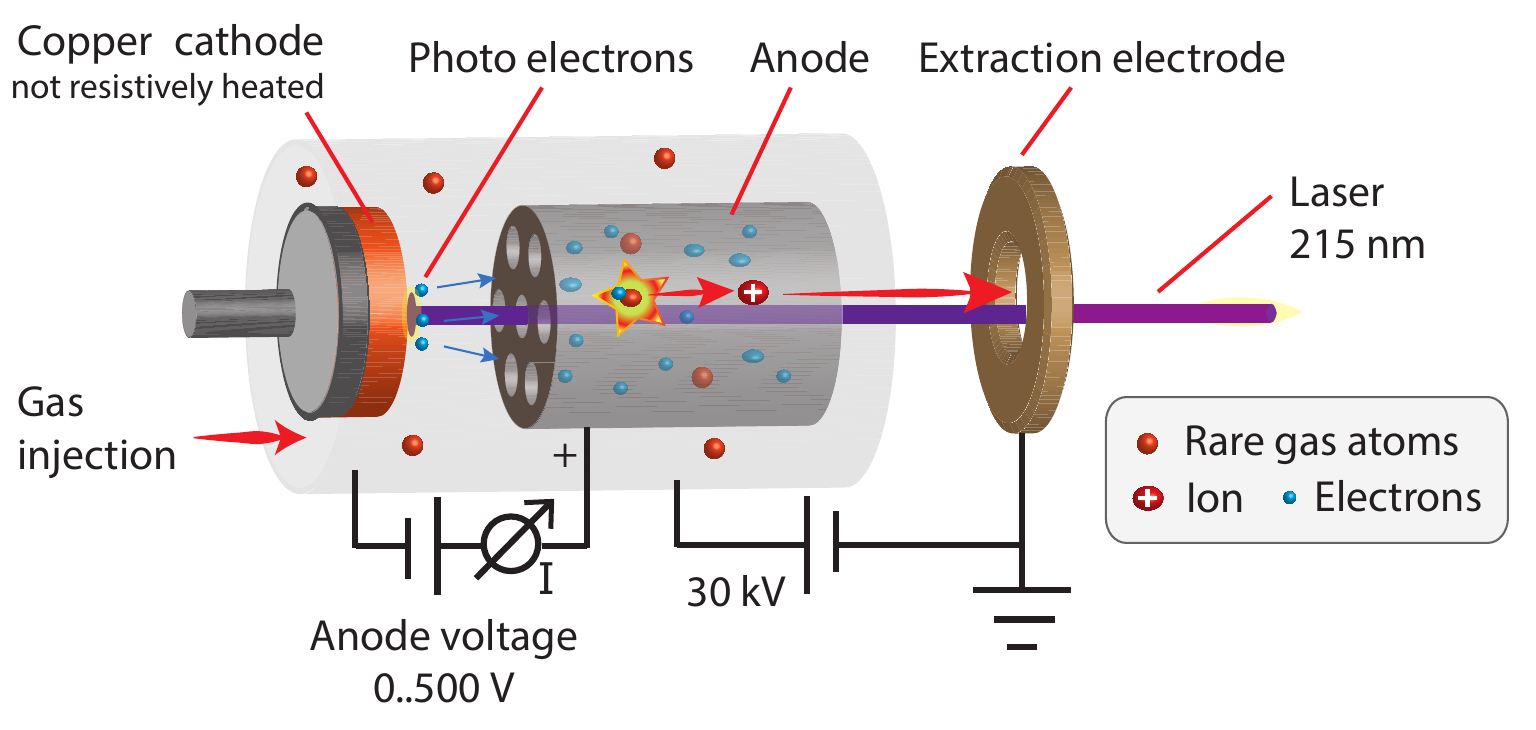}
  \vspace{0.01cm}
}{%
  \caption{\label{fig:CuVADISschema} Schematic of the electron-impact ion source used in our experiment. The electrons were liberated from a copper cathode by a laser beam (photo-electric effect) in a cold environment.}%
}
%\capbtabbox{%
\killfloatstyle
\ttabbox[\FBwidth]{%
   	\footnotesize
 \begin{tabular}{l|r|l}
	parameter &  value & unit\\
	\hline
	\multicolumn{3}{l}{} \\
	\multicolumn{3}{l}{ion source properties} \\
	\hline
	anode-cathode gap & \num{2.8\pm 0.2} & \si{\milli\meter} \\
	cathode material & Cu &  \\
	\multicolumn{3}{l}{} \\
	\multicolumn{3}{l}{laser properties (on the cathode)} \\
	\hline
	center wavelength & \num{215} & \si{\nano\meter} \\
	repetition rate & 10 & \si{\kilo\hertz} \\
	beam diameter & \num{1.5} & \si{\milli\meter}  \\
	average power & $\sim \num{10}$ & \si{\milli\watt}  \\
	pulse length & $\sim \num{30}$ & \si{\nano\second}  \\
	pulse energy & $ \sim \num{1}$ & \si{\micro\joule}  \\
	pulse energy fluence & $\sim \num{56}$ & \si{\micro\joule\per\square\centi\meter}  \\
\end{tabular}

}{%
  \caption{Parameters of the photo-cathode experiment.}%
}
\end{floatrow}
\end{figure}

\section{Experimental setup}

The experiments were conducted at the ISOLDE Offline 2 separator \cite{Warren2020}. The ion source used within this work is based on the VADIS version VD7 equipped with a water-cooled transfer-line \cite{Penescu2010,LiviuDiss}. The VADIS VD7 is often referred to, quite incorrectly, as cold Forced Electron Beam Induced Arc Discharge (FEBIAD) ion source. It is schematically shown in fig.~\ref{fig:CuVADISschema}. The ion source is part of a Target and Ion Source (TIS) unit that is equipped with a coil that generates a longitudinal magnetic field (in the following referred to as target magnet). The coil can be supplied with currents of up to ca. \SI{6}{\ampere} which are expected to produce a magnetic field of up to ca.~\SI{30}{\milli\tesla} \cite{MartinezPalenzuela:2672954}. The anode was biased up to \SI{500}{\volt}.

The photo-cathode was built by pressing a copper cap on the tantalum cathode of a standard VADIS. The cap was manufactured from a sheet of polished oxygen-free copper. It was degreased in pentane and oxide layers were removed by an aqueous citric acid solution. The cap was rinsed with demineralized water and finally subjected to ultra-sound cleaning in ethanol. The cathode was electrically insulated from the target assembly to enable measurements of the photo-electron current. Additionally, the electron current was obtained as drain current of the anode power-supply. The anode drain current measurement does not differentiate between electrons impinging on the anode grid and electrons entering the anode volume. Differences between cathode and anode drain current might arise \textit{e.g.}, due to the fraction of electrons emerging from the cathode but not arriving at the anode due to space-charge repulsion, or by electron generation by the laser beam on the ion extraction side where metal disks (typically serving as heat shields) are located that are electrically insulated from the anode. Electrons produced in this region are repelled by the extraction electrode and accelerated towards the anode.

 The cathode was not baked prior to, or during operation. A deep-ultraviolet laser beam (\SI{215}{\nano\meter}, corresponding to \SI{5.77}{\electronvolt} photon energy) was guided through the ion source outlet orifice, through the central hole of the anode grid (both \SI{1.5}{\milli\meter} in diameter), and onto the cathode surface.  The UltraViolet (UV) beam was generated by frequency quadrupling of a Ti:sapphire laser, identical to the ISOLDE resonance ionization laser ion source setup \cite{Fedosseev_2017}. It provided \SI{30}{\nano\second}-long pulses at a repetition rate of \SI{10}{\kilo\hertz}. Prior to installation of the ion source on the beam line, the laser beam transmission through the mass separator window and ion source aperture was measured with a laser alignment device that resembles the geometry of a VADIS. The laser beam was focused such that ca.~\SI{73}{\percent} of the power arriving at the ion source reached the cathode surface. The remaining fraction was stopped by the aforementioned heat shields. During the experiment, the laser power was measured in front of the vacuum window. The transmission between the latter position and the cathode was measured to be \SI{35 \pm 2}{\percent}. The power arriving at the cathode was ca.~\SI{10}{\milli\watt} and was varied during the experiment. Krypton (Carbagas, 99.998\%) and carbon dioxide (Carbagas, 99.998\%) at a pressure of \SI{1.2}{\bar} were supplied through a calibrated leak of \SI{1.7e-5}{\milli\bar\liter\per\second} (\SI{1}{bar} abs., air). In addition to Faraday cups, the ion beam was assessed with a MagneToF detector (ETP DM291). The latter was used to determine the time-structure of ion release. The ion beam transmission through the mass separator was estimated to be \SI{87}{\percent}.

\section{Results and discussion}

The previously described setup was operated under varying conditions for about 6 days. The laser was continuously irradiating the cathode during this period. During the first days, krypton was supplied though the calibrated leak. For the last 66 hours, carbon dioxide was injected. The ion and electron currents were measured as a function of anode voltage, magnet current and laser power. 

%Bruce: im not sure if this statement is needed if you have electron current measurements .  it is too vague to be useful to the reader: 
%The laser system was typically optimized at least once per day due to instabilities of position and beam shape, leading to a significant drop of laser power during each day. 

%when the target assembly was biased to \SI{30}{\kilo\volt}
%(\SI{100}{\volt} anode voltage, \SI{5.5}{\ampere} magnet current)
%Loogbook written by ofconso @pcen36897.cern.ch (2021/08/29 16:52:55)
The photo-electron currents were measured as cathode current and anode drain current. Both were in agreement within ca. \SIrange{10}{20}{\percent}, while the anode drain current was typically higher. 
The maximum observed \textsuperscript{84}Kr current at a relatively strong magnetic field (\SI{5.5}{\ampere}) was ca.~\SI{1}{\nano\ampere}, which corresponds to an ionization efficiency of ca.~\SI{0.004}{\percent}.
At the same time, the anode drain current was measured to be ca. \SI{100}{\nano\ampere}. At the same anode voltage of \SI{100}{\volt} and without magnet field, the electron current increased to ca. \SI{270}{\nano\ampere} which is in agreement with the expected space-charge limited current of $\SI[parse-numbers=false]{263^{+42}_{-37}}{\nano\ampere}$ as given by the Child-Langmuir equation \cite{Child1911,Langmuir1913} for a one dimensional emitter, and an anode-cathode distance of \SI{2.8 \pm 0.2}{\milli\meter}. 
%
% Quantum efficiency from power scan
% error based on: 5% transmission error + power error (0.3 + 5%) + 10% current error
The highest drain current of ca. \SI{1.5}{\micro\ampere} was obtained at an anode voltage of \SI{500}{\volt} and without magnetic field.  The space-charge limit at \SI{500}{\volt} is expected to be ca.~\SI{2.5}{\micro\ampere}. The quantum efficiency computes to~\num[exponent-mode = fixed, fixed-exponent = -4]{0.0008 \pm 0.0002}. The value exceeds efficiencies reported by other authors by a factor of ca.~\num{2} \cite{palmer2005review}, and could indicate that the laser beam transmission during the experiment exceeded the estimate of $\SI{35\pm2}{\percent}$ that was obtained during alignment.

\begin{figure} [t]
\begin{minipage}{18pc}
\includegraphics[width=18pc]{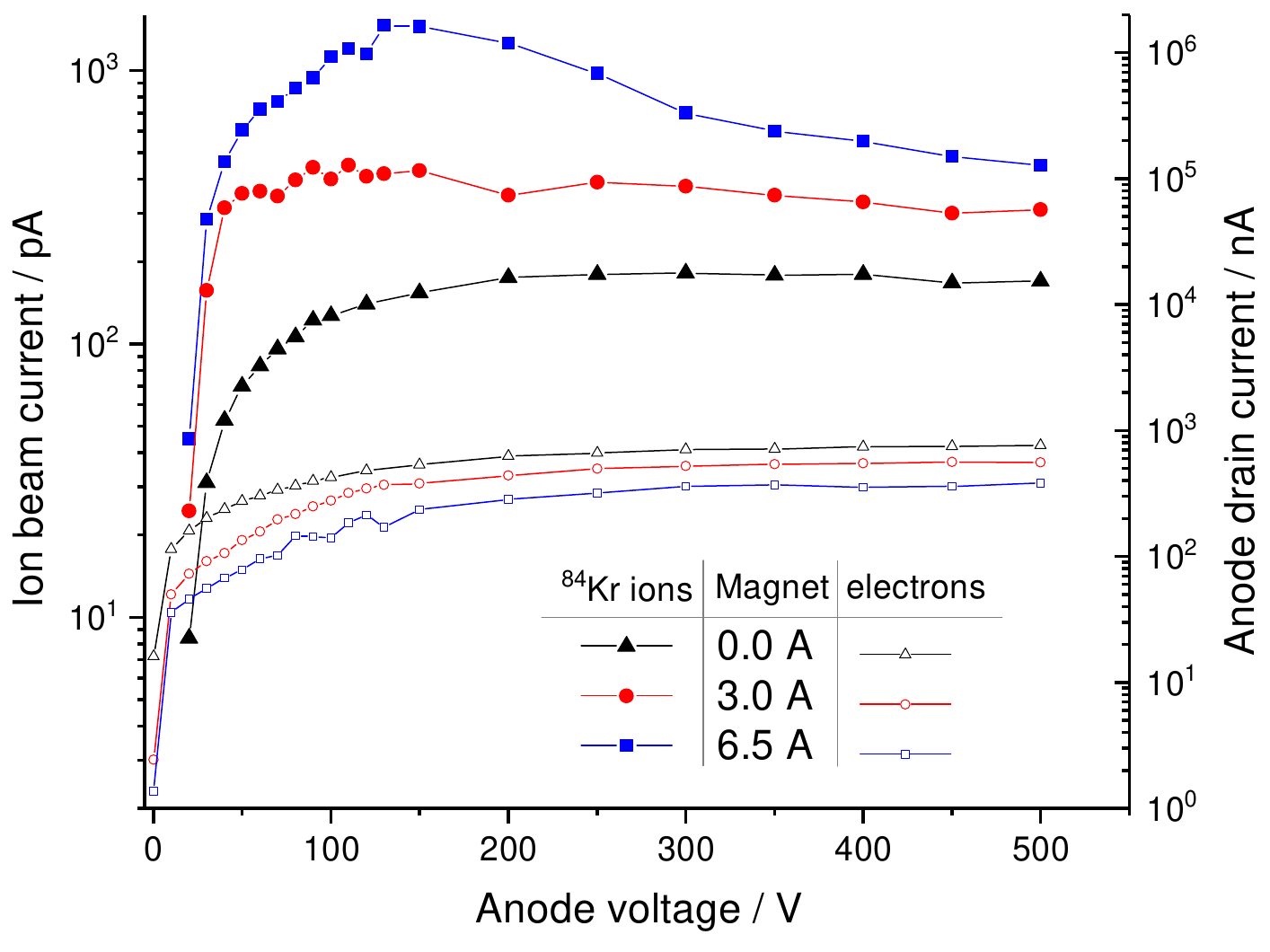}
\caption{\label{fig:anodescan} (left) Dependence of anode drain current and \textsuperscript{84}Kr ion current on the anode voltage for different settings of the target magnet at a laser power and center wavelength of \SI{8(1)}{\milli\watt} and \SI{215}{\nano\meter}, respectively. Connecting lines were added to guide the eye. (right) Time structure of electron, laser and ion pulses. A photo-diode was used to assess the laser time structure. Laser and electron pulse shape were estimated by measurement of the voltage drop over a \SI{50}{\ohm} resistor. The curve shapes were redrawn from an oscilloscope photograph. The offset between electron and laser pulses was not measured. The ion time structure was obtained with a MagneToF detector and is given relative to the laser pulse.}
\end{minipage}\hspace{2pc}%
\begin{minipage}{16pc}
\includegraphics[width=16pc]{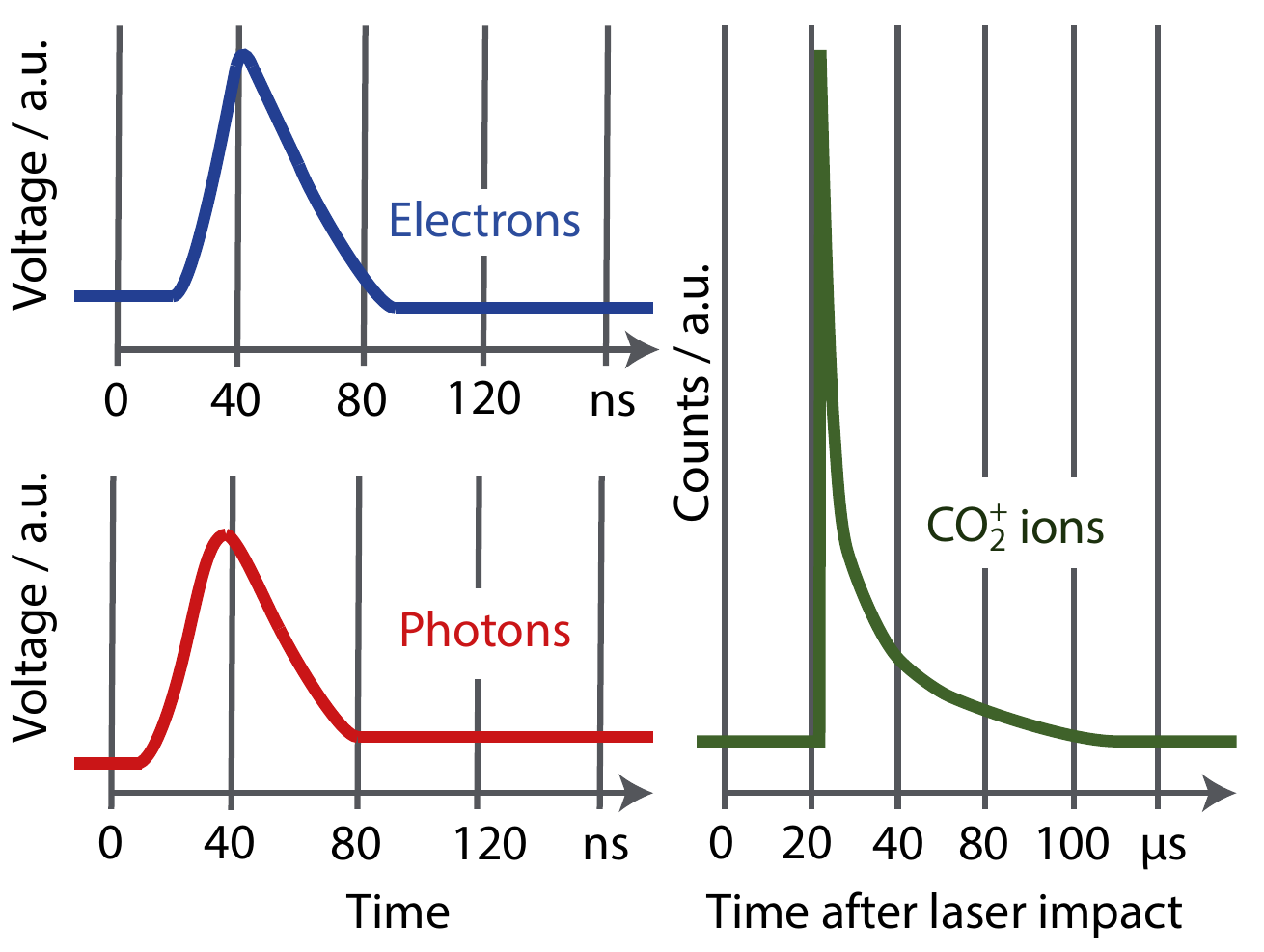}
%\caption{\label{fig:powerscanions} ion current current vs. laser power}
\end{minipage} 
\end{figure}

The evolution of ion and anode drain current while varying the anode voltage (anode scan) is given for three settings of the target magnet in fig.~\ref{fig:anodescan}. The target magnet significantly increases the ion currents and decreases the electron currents. This effect is typically explained by an increase of electron density in the ionization volume, which in turn increases the ionization efficiency \cite{LiviuDiss}. At the same time, the higher charge density might reduce the total electron emission due to space-charge repulsion. Without magnetic field, the ion and electron currents increased steadily with anode voltage, likely due to a combination of lower space-charge repulsion at higher voltages and a focusing effect of the electrostatic field. The plateau region (\SIrange{250}{500}{\volt}) could indicate that the electron supply is rather limited by quantum efficiency than space-charge repulsion. At elevated magnet currents (\SI{6.5}{\ampere}), a pronounced maximum in the ion current emerges at ca. \SI{150}{\volt} anode voltage, which can be explained by the electron-energy dependence of electron impact ionization cross sections \cite{Higgins:203081}.

\begin{figure} [t]
\begin{minipage}{17pc}
\includegraphics[width=17pc]{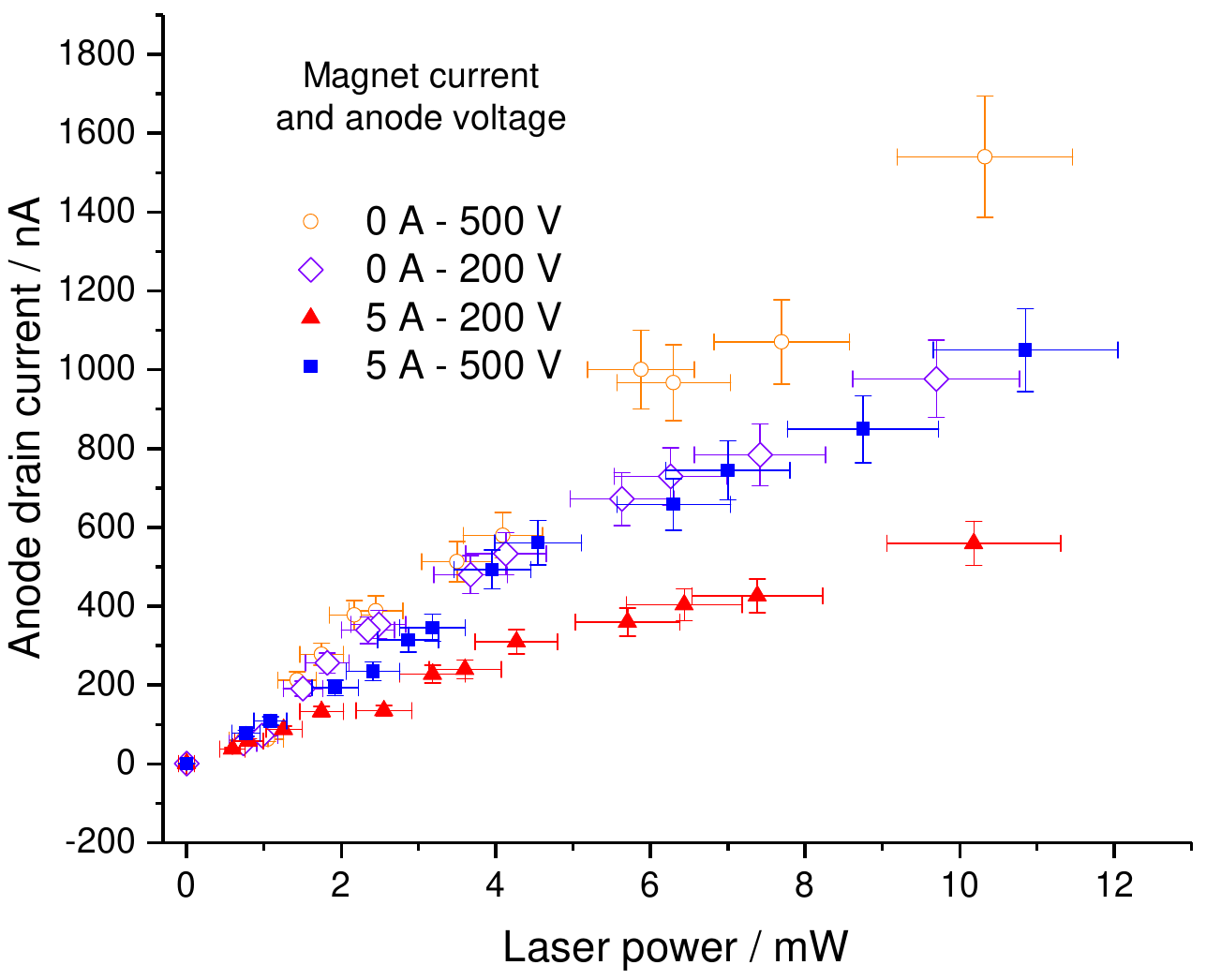}
\caption{\label{fig:powerscan} Dependence of anode drain current and \textsuperscript{84}Kr ion current on laser power for different anode voltages and magnet currents. The electron current measurement does not differentiate between electrons impinging on the grid and electrons entering the anode volume. The ion currents for \SI{500}{\volt} and \SI{200}{\volt} at \SI{0}{\ampere} magnet current were indistinguishable, and only the data for \SI{500}{\volt} are shown.
%Bruce:
% if there is a electron current dependency on the % of electrons reaching the anode volume, this could account for the different slopes for the LH and RH plots.
}
\end{minipage}\hspace{2pc}%
\begin{minipage}{17pc}
\includegraphics[width=17pc]{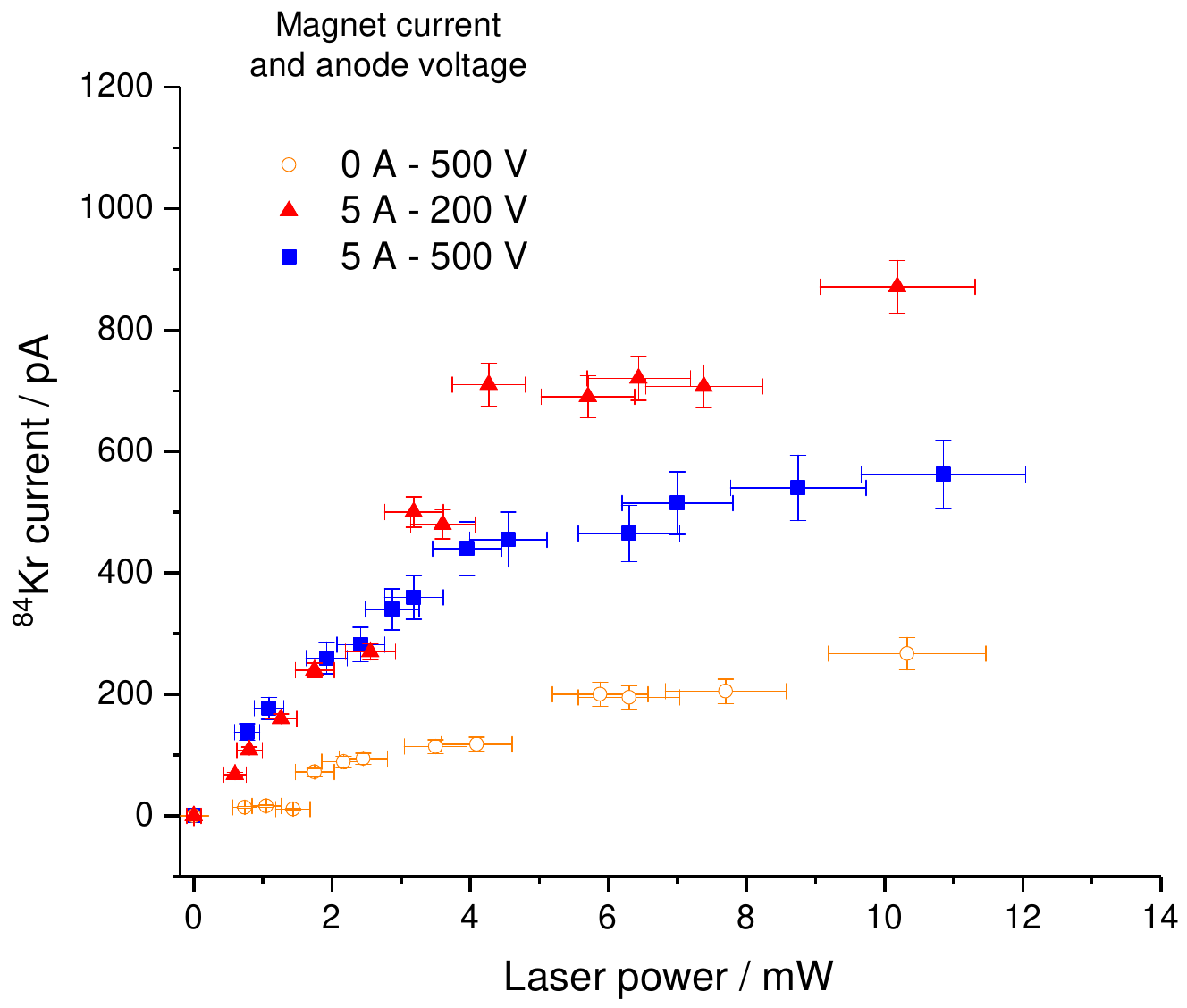}
%\caption{\label{fig:powerscanions} ion current current vs. laser power}
\end{minipage} 
\end{figure}

The dependence of ion beam current and anode drain current on laser power is shown in fig.~\ref{fig:powerscan}. The measurements were suffering from relatively high errors due to instabilities of ion beams, data acquisition and laser power. Ion and electron currents steadily increased with laser power. The ion currents, in particular the data for a magnet current of \SI{5}{\ampere} and \SI{500}{\volt} anode voltage,  show saturation, \textit{i.e.}, an increase in laser power and electron production does not translate into a proportional increase in the ion beam intensity. It seems likely that despite increased electron production, electrons either do not reach the anode volume, their energy is not sufficient for ionization or the probability of ion extraction is negatively impacted.

\begin{figure} [t]
\includegraphics[width=38pc]{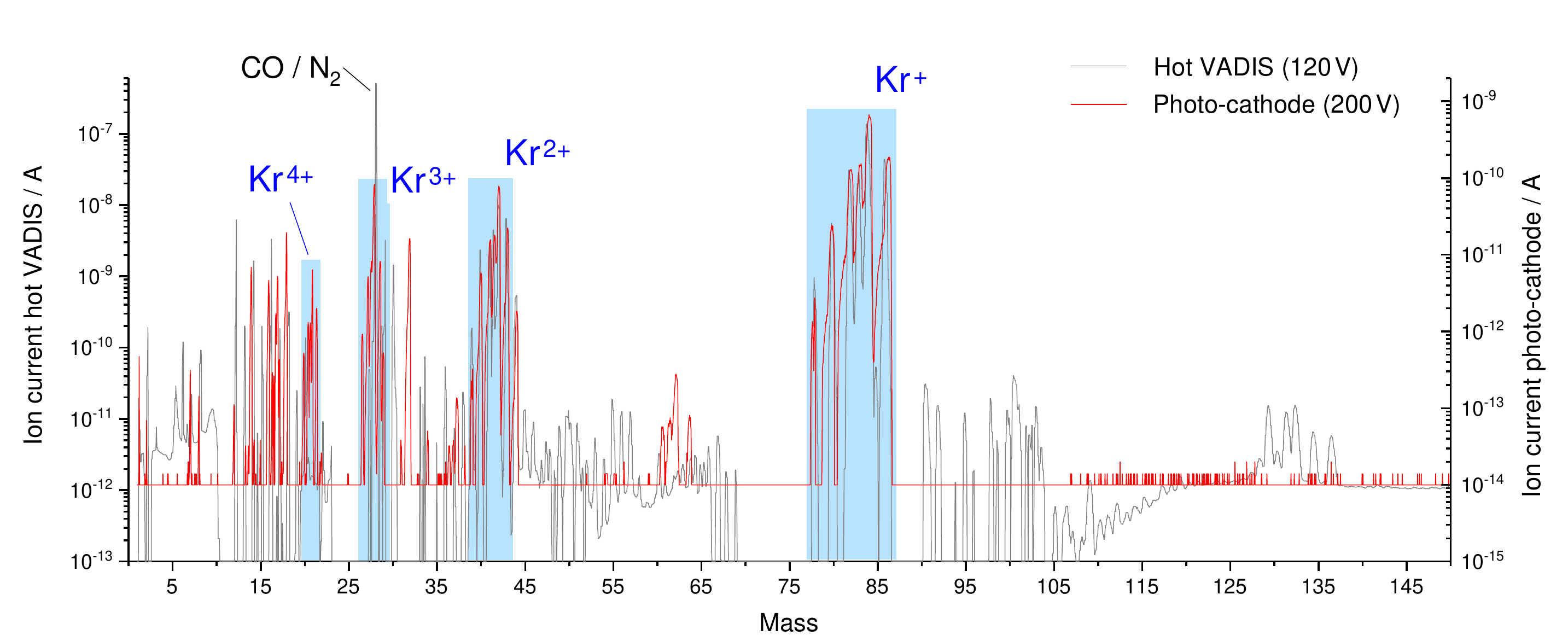}
\caption{\label{fig:mass-scan} Mass scan obtained with the photo-cathode ion source obtained at an anode voltage of \SI{200}{\volt} in comparison to a mass scan obtained with a standard VADIS (VD7, thermionic electron emission) operated at \SI{120}{\volt}. Both sources were supplied with krypton. In addition, the standard VADIS was supplied with carbon monoxide gas and residual xenon was present from previous tests. The scales have been adjusted to show a similar level of sensitivity. Negative ion currents for the photo-cathode source have been trimmed to \SI{1e-14}{\ampere}.}
\end{figure}

A mass spectrum obtained with the photo-cathode driven ion source in comparison to a regular VADIS is shown in fig.~\ref{fig:mass-scan}. Due to operation at ambient temperature the level of impurity is significantly lower for the photo-cathode ion source.

The quantum efficiency of a photo-cathode driven ion source could be impacted by poisoning of the cathode surface due to condensation of molecule fragments produced in electron collisions and the unusually high rest gas pressure for photo-cathode operation \cite{palmer2005review}. 
To evaluate the photo-cathode lifetime in these conditions, we have injected carbon dioxide into the ion source for ca.~$66\;\mathrm{hours}$. The ionization efficiency for the extraction as $\mathrm{CO^{+}_{2}}$ was measured to be ca. \SI{2e-5} .. An immediate effect on the anode drain current could not be observed. However, after $66\;\mathrm{hours}$, the anode drain current reduced to ca.~\SI{35}{\percent} of its original value. The gas pressure inside the ion source was estimated to be ca. \SI{1e-4}{\milli\bar}, following the calculation proposed in \cite{LiviuDiss}, and under consideration of the gas flow rate through the calibrated leak and the conductance of the outlet orifice. The typical pressure inside the ion source of a hot VADIS is far below this value and in the order of \SI{e-6}{\milli\bar}.

% blue light laser power 440 mW * 0.35% transmission = 154 mW

Measurements of the electron, photon and ion beam time structure are provided in fig.~\ref{fig:anodescan}. The electron pulse shape closely resembles the laser pulse shape as measured by a photo-diode. The $\mathrm{CO}_{2}^{+}$ ions are, in contrast to what has been observed for resonantly ionized atomic species inside a hot VADIS source \cite{Goodacre2016}, contained in a single bunch and arrive ca. \SI{20}{\micro\second} after the laser impact on the cathode at the MagneToF detector. This difference supports the assumption that, for this study, the VADIS is acting predominantly as an electron impact ion source without ion and electron dynamics that govern ion survival and extraction in standard operating conditions of the VADIS.

To confirm that the electron production was governed by the photo-electric effect, blue light (\SI{430}{\nano\meter}, corresponding to \SI{2.88}{\electronvolt} photon energy) at a power of \SI{154}{\milli\watt} (second harmonic generation from the same laser system) was guided on the cathode. The work function of copper (\SIrange{4.5}{4.9}{\electronvolt} \cite{CRC2020}) is above the photon energy at that wavelength, so that a single-photon induced electron emission is not possible.  In agreement with expectations, no electron production could be observed.

%Reinhard:
%I did a quick calculation here: I get ~22us travel time at 30kV to the %detector, that seems to fit. As expected, that would mean we see mainly %ions that are created directly at the VADIS exit, but also a tailing of %ions that first have to travel inside the source to the exit. So there %seems to be some kind of confinement - probably not thermal movement of %Kr into the extraction, that is only ~ 0.25 mm/us at room temperature. %Yet there is not the pronounced second peak of ions created directly at %the anode as with VADLIS.

%That is not the scope of this paper though, but we can discuss at some %point what we learn from this.

% Scaling for reduced electron current:
%5A 130V ?
%
%from anode scan:
%Current 200V: 285 nA
%Current 130V: 171 nA
%----> factor 0.6 scanling 200V -> 130V anode current
%
%from power scan:
%5A 200V = 560 nA 
%with scaling factor:
%5A 130V = 336 nA
%
%66 hours of CO2: 117 nA
%
%--> 35% left.

% With an estimated laser power on the cathode of  

%- Anode Scan, magnet off, ofconso @pcen36897.cern.ch (2021/09/01 14:28:25)
%- Magnet scan, Anode at 500, and as before.  written by ofconso @pcen36897.cern.ch (2021/09/01 14:27:46)

\section{Conclusions and outlook}

%Do you want to say anything about the importance of post-breakup/ionization ion confinement that is required for non volatile ions?  This would have to ba a key feature of any final application of the PC source for refractory metals, and would surely be factor contributing to a  discrepancy between observed effciency for volatile species such as Kr, and e.g Mo

In this study we have demonstrated production of molecular ion beams using an electron impact ion source driven by a photo-cathode at ambient temperature. In combination with high in-target production rates of radioisotopes, the existing ion source design would in principle allow first experiments with fragile radioactive molecules. We have previously proposed a parameter study to increase the ion source efficiency to ca. \SI{1}{\percent}, which would allow a wide range of experiments with exotic molecules \cite{ballof2021concept}. A major efficiency increase requires the modification of the ion source geometry to resolve the space-charge limitation implied by the \SI{1.5}{\milli\meter} laser beam spot on the cathode surface. A further efficiency increase, especially for molecules disintegrating to non-volatile species, could likely be obtained by embedding the anode volume in an ion guiding structure, \textit{e.g.}, an electric quadrupole as applied in the Laser Ion Source and Trap (LIST) ion source \cite{Fink2013}. 

The new ion source will enable extraction and ionization of molecules at ISOLDE which are not stable at ultra-high temperatures. Development and implementation of this method at ISOLDE will expand its radioactive ion beam capabilities to refractory elements \cite{ballof2021concept} and help to serve the increasing demand in radioactive molecules for spectroscopy applications \cite{Ruiz2020,Safronova2018,Athanasakis:2748712}.

\ack

We would like to thank Mathieu Bovigny and Julien Thiboud for the assembly of the target unit used in our experiments, James Cruikshank for the support regarding the mass separator and Eric Chevallay for sharing his expertise about copper photo-cathodes. This project has received funding from the European Union’s Horizon 2020 research and innovation program under grant agreement Nos. 654002 and 861198–LISA–H2020-MSCA-ITN-2019.

\bibliography{biblio}

\end{document}